# Extended quasi-additivity of Tsallis entropies

Ryszard Piasecki[*]

*Institute of Chemistry, University of Opole, Oleska 48, 45-052 Opole, Poland*

**Abstract**

We consider statistically independent *non-identical* subsystems with different entropic indices $q_1$ and $q_2$. A relation between $q_1$, $q_2$ and $q'$ (for the entire system) extends a power law for entropic index as a function of distance $r$. A few examples illustrate a role of the proposed constraint $q' < \min(q_1, q_2)$ for the Beck's concept of quasi-additivity.

*Keywords:* Tsallis statistics; Quasi-additivity

## 1. Introduction

The description of nonextensive systems within Tsallis statistics [1] (see, e.g., [2] for recent applications) is based on postulated non-additive $q$-entropy

$$S_q = \frac{1}{q-1}\left[1 - \sum_i^W p_i^q\right], \qquad (1)$$

where $W$ is the number of accessible microscopic states of a system under consideration with the associated probabilities $\{p_i\}$ and $q$ is positive real number termed the entropic index related to the degree of nonextensivity. Tsallis' original suggestion was that this approach may be relevant for equilibrium systems with long-range interaction, but recently it was pointed out that the formalism comes into play when systems are far from equilibrium. For example, for systems with fluctuating mean free path [3] (for $q > 1$) and [4] (for $q < 1$), temperature or energy dissipation rate [5, 6] (for $q > 1$). The common feature of the mentioned examples of non-equilibrium systems is that the parameter $q$ can be expressed by the relative variance of the fluctuations of a parameter $X$

$$q = 1 \pm \frac{\langle X^2 \rangle - \langle X \rangle^2}{\langle X \rangle^2} \qquad (2)$$

provided $X$ is $\chi^2$ (or gamma) distributed. Here, the symbols '+' ('−') refer to the $q > 1$ ($q < 1$) cases, respectively. From this point of view, the parameter $q$ can be treated in such systems as a measure of the fluctuations of a parameter $X$ [3, 4].

---

[*]Fax: +48 77 4410740.
 *E-mail address:* piaser@uni.opole.pl (R. Piasecki).

Recently, Beck [7] has suggested that for systems with fluctuations in temperature or energy dissipation rate it is possible to make the Tsallis entropies quasi-additive by choosing different entropic indices at different spatial scales: $q$ for statistically independent *identical* subsystems I and II and $q' < q$ for composed system I + II. For systems with locally fluctuating inverse temperature $\beta$ we have $q > 1$ and if they are turbulence systems then $q$ usually is close to one. Moreover, there are experimental examples showing that $q$ is a monotonously decreasing as a function of distance $r$ [8]. According to Beck [7], such scale dependence is a result of the following quasi-additivity property of the Tsallis entropies for $q'$ properly chosen

$$S_q^I + S_q^{II} = S_{q'}^{I+II}, \qquad (3)$$

where statistically independent *identical* subsystems appear. For such subsystems the number of their microstates as well as the type of microstate probability distributions are assumed to be the same, that is $W_I = W_{II}$ and $\{p_i(W_I)\} = \{p_j(W_{II})\}$. Then, from Eq. (3) for $q$ close to 1 the following relation between $q$ and $q'$ can be obtained [7]

$$q' - 1 = (q - 1) \frac{\langle B_i^2 \rangle}{\langle B_i^2 \rangle + \langle B_i \rangle^2} \quad \Rightarrow \quad q' < q, \qquad (4)$$

where $B_i := \log p_i$ is so called ''bit number'' [9], its negative expectation $-\langle B_i \rangle := -\Sigma p_i \log p_i$ is the Shannon entropy and variance $\langle B_i^2 \rangle := \Sigma p_i \log^2 p_i$ is related to fluctuations of the entropy.

Very recently has been reported another interesting concept [10–12]. Within this approach equal and distinguishable subsystems can be strongly (globally) *correlated* such that for an adequate value of $q \neq 1$ the $S_q$ becomes strictly additive. This means that under specially correlated composition of subsystems even for entropic index the same for composed system and subsystems (now they are not statistically independent) it is still possible to make $S_q$ additive quantity.

Let us clarify the discussion by denoting: (A) = subsystems are independent, (B) = $q$ is the same for composed system and subsystems, and (C) = $S_q$ is additive. Within this language most early papers on nonextensive statistical mechanics keep (A) and (B) and give up (C), Beck (in Ref. [7]) keeps (A) and (C) and gives up (B) while Tsallis, Gell-Mann and Sato (in Ref. [12]) keep (B) and (C) and give up (A). Therefore, the completion of the Beck's paper by including the most general case, i.e. (A') = *non-identical* subsystems with different entropic indices $q_1$ and $q_2$ are independent, is a natural motivation for the present study.

## 2. Extended case

The present work concerns the extension of the quasi-additivity property in Eq. (3) and the generalization of relation given by Eq. (4) to the systems composed of two statistically independent but *non-identical* subsystems by choosing different entropic indices $q_1$ and $q_2$ at different system's size. We shall propose the extended relation

$$S_{q_1}^I + S_{q_2}^{II} = S_{q'}^{I+II}. \tag{5}$$

From now on, we refer the subscript $i$ ($j$) to subsystem I (II), respectively. For the entire system we use the following rule for probability distributions, $\{p_{ij}(W_{I+II})\} = \{p_i(W_I)\, p_j(W_{II})\}$. Now the extended linear relation between $q_1$, $q_2$ and $q'$ can be obtained in a similar way as described in Ref. [7]. To make our paper self-contained we present the basic steps of this procedure. Assuming that both entropic indices $q_1$ and $q_2$ are close to 1 and using Taylor expansion in $q_1 - 1$ and $q_2 - 1$ for Eq. (5) one obtains

$$\begin{aligned} S_{q_1}^I + S_{q_2}^{II} &= -\sum_i^{W_I} p_i \ln p_i - \sum_j^{W_{II}} p_j \ln p_j - \frac{1}{2}(q_1-1)\sum_i^{W_I} p_i \ln^2 p_i \\ &\quad - \frac{1}{2}(q_2-1)\sum_j^{W_{II}} p_j \ln^2 p_j - \ldots \\ &= -\langle B_i \rangle - \langle B_j \rangle - \frac{1}{2}(q_1-1)\langle B_i^2 \rangle - \frac{1}{2}(q_2-1)\langle B_j^2 \rangle - \ldots \end{aligned} \tag{6}$$

and similarly,

$$\begin{aligned} S_{q'}^{I+II} &= \frac{1}{q'-1}\left[1 - \left(\sum_i^{W_I} p_i^{q'}\right)\left(\sum_j^{W_{II}} p_j^{q'}\right)\right] \\ &= -\langle B_i \rangle - \langle B_j \rangle - \frac{1}{2}(q'-1)[\langle B_i^2 \rangle + \langle B_j^2 \rangle + 2\langle B_i \rangle\langle B_j \rangle] - \ldots \end{aligned} \tag{7}$$

Equating the coefficients of same degree and neglecting higher-order contributions in $q_1 - 1$, $q_2 - 1$ and $q' - 1$ we obtain the following relation[†]

$$q' - 1 = \frac{(q_1-1)\langle B_i^2 \rangle + (q_2-1)\langle B_j^2 \rangle}{\langle B_i^2 \rangle + \langle B_j^2 \rangle + 2\langle B_i \rangle\langle B_j \rangle} \quad \Rightarrow \quad q' < \omega q_1 + (1-\omega)q_2, \tag{8}$$

where $\omega = \langle B_i^2 \rangle/(\langle B_i^2 \rangle + \langle B_j^2 \rangle)$. Since for equally sized identical subsystems $q_1 = q_2 \equiv q$, $W_I = W_{II}$ and $\{p_i(W_I)\} = \{p_j(W_{II})\}$, then we have $\langle B_i \rangle = \langle B_j \rangle$, $\langle B_i^2 \rangle = \langle B_j^2 \rangle$, thus $\omega \equiv 1/2$ and Beck's result given by Eq. (4) is recovered. The corresponding to Eq. (8) a non-linear relation between $q_1$, $q_2$ and $q'$ that includes terms up to the second order in $q_1 - 1$, $q_2 - 1$ and $q' - 1$ is shown in Appendix.

One point more is worth to be considered. The quasi-additivity property for systems with fluctuations necessarily implies that entropic index $q(r)$ is a strictly monotonously decreasing function of distance $r$ [7]. Exactly the same arguments as those used by Beck in Ref. [7] on page 331, can be applied to our case of three systems different in sizes: I, II, and I + II. We postulate that *physically* admissible $q'$, $q_1$ and $q_2$ values should satisfy a general constraint

---

[†] Eq. (8) was discussed briefly with Prof. C. Beck on NEXT 2001 (Cagliari)

$$q' < \min(q_1, q_2). \tag{9}$$

For identical subsystems this constraint becomes equivalent to Beck's one already included in Eq. (4). However, for non-identical subsystems, as shown later, certain pairs of model entropic indices $q_1$ and $q_2$ can violate the above constraint. Such combinations of the indices are not allowed from the physical point of view.

Finally, a simple scaling law considered in Ref. [7] and connected with a transformation of distance $r \to \lambda r$ can be also extended. Without any loss of generality let us consider $1 < q_1 \leq q_2$. Denoting $\alpha \equiv (q_2 - 1)/(q_1 - 1)$, Eq. (8) can be rewritten as

$$q' - 1 = (q_1 - 1)(\Gamma_1 + \alpha \Gamma_2), \tag{10}$$

where $\Gamma_1 = \langle B_i^2 \rangle / D$, $\Gamma_2 = \langle B_j^2 \rangle / D$ and $D = \langle B_i^2 \rangle + \langle B_j^2 \rangle + 2 \langle B_i \rangle \langle B_j \rangle$. If $q_1$ and $q_2$ are close to 1 (or $r$ larger enough), the functional coefficients $\alpha$, $\Gamma_1$ and $\Gamma_2$ are approximately constant and Eq. (10) can be now iterated. Similarly to Ref. [7], after $n$ steps one obtains $r = \lambda^n r_0$ and

$$\frac{1}{q-1} = \frac{1}{q_0 - 1} (\Gamma_1 + \alpha \Gamma_2)^{-n}. \tag{11}$$

Writing Eq. (11) in the form

$$\frac{1}{q-1} = \frac{1}{q_0 - 1} \left(\frac{r}{r_0}\right)^{\Delta}, \tag{12}$$

eliminating $n$ and comparing r.h.s. of Eqs. (11-12) the scaling index can be written as

$$\Delta = -\log_\lambda (\Gamma_1 + \alpha \Gamma_2). \tag{13}$$

Notice that for identical subsystems we have $\Gamma_1 + \alpha \Gamma_2 \to b$ and $\Delta \to \delta = -\log_\lambda b$, where $b := \langle B_i^2 \rangle / (\langle B_i^2 \rangle + \langle B_i \rangle^2) \equiv \langle B_j^2 \rangle / (\langle B_j^2 \rangle + \langle B_j \rangle^2)$. Thus for $\lambda = 2$ one recovers Beck's result [7]. A scaling law in $r$ of type (12) (in Beck's version for identical subsystems) was observed in Couette-Taylor experiment of Swinney *et al.* [8] within the exponent range $\delta \approx 0.33$–$0.44$. The Swinney data are also compatible with a power law for $1/(q-1)$ as a function of $r$ predicted by Beck within quasi-additivity approach. The observed exponent was $\delta \approx 0.30$ (see Fig. 4 in [13]). On the other hand, in the Gaussian approximation for turbulence application, a rough estimate of the exponent $\delta$ leads to $\delta \approx 0.42$ (see Eq. (20) in [7].

## 3. Examples

To investigate the functional dependence $q' \approx q'(q_1, q_2)$ we use in Eq. (14) so called $q$-exponential, first considered in the nonextensive context by Tsallis [1]. With respect to turbulent flows [14], where the kinetic energy $u^2/2$ is associated with the radial velocity difference $u$ between two points in the liquid separated by a distance $r$, the probability density $p(u)$ after extremizing the Tsallis entropy $S_q$ is given by

$$p(u) = \frac{1}{Z_q}[1 + \frac{1}{2}(q-1)\beta u^2]^{-1/(q-1)} . \qquad (14)$$

Here $Z_q$ is the partition function existing under condition that $1 \leq q < 3$ and for $k := 1/(q-1) \geq 2$ being an integer the simple form of $Z_q$ can be obtained [14]

$$Z_q = \int_{-\infty}^{+\infty} p(u)du = \sqrt{\frac{2k}{\beta}} \frac{\pi(2k-3)!!}{(k-1)!2^{k-1}} . \qquad (15)$$

Now, applying the extended quasi-additivity property (5) and resulting linear relation (8) in its continuous version one evaluates explicitly [15]

$$<\log p(u)> = k\left[\Psi\left(k-\frac{1}{2}\right) - \Psi(k)\right] - \log Z_q \qquad (16)$$

and

$$<\log^2 p(u)> = k^2\left\{\left[\Psi\left(k-\frac{1}{2}\right) - \Psi(k)\right]^2 + \Psi'\left(k-\frac{1}{2}\right) - \Psi'(k)\right\}$$
$$- 2k\left[\Psi\left(k-\frac{1}{2}\right) - \Psi(k)\right]\log Z_q + \log^2 Z_q , \qquad (17)$$

where digamma function

$$\Psi(k) = -\gamma + \sum_{i=1}^{k-1}\left(\frac{1}{i}\right), \qquad (18a)$$

$$\Psi(k-\frac{1}{2}) = -2\ln 2 - \gamma + \sum_{i=1}^{k-1}\left(\frac{2}{2i-1}\right), \qquad (18b)$$

$$\Psi'(k) = \pi^2/6 - \sum_{i=1}^{k-1}\left(\frac{1}{i^2}\right), \qquad (18c)$$

$$\Psi'(k-\frac{1}{2}) = \pi^2/2 - \sum_{i=1}^{k-1}\left(\frac{4}{(2i-1)^2}\right) \qquad (18d)$$

and $\gamma \cong 0{,}577216$ is Euler's constant. For $\beta = 2/(5-3q) = 2k/(2k-3)$ the probability density $p(u)$ has average value 0 and variance 1. To plot the functional dependence $q' \approx q'(k_1(q_1), k_2(q_2)) = q'(1/(q_1-1), 1/(q_2-1))$ according to Eq. (8), we consider all pairs of chosen integers $k_1, k_2 \in \{3, ..., 15\}$ for two constant values of $\beta$, see Fig. 1a with $\beta = 2.0$ and Fig. 1b with $\beta = 1.12$. So, both subsystems with corresponding $q_1 := 1 + 1/k_1$ and $q_2 := 1 + 1/k_2$ indices have common temperature but the subsystem's probability densities slightly differ in variances.

Unexpectedly, only the sites marked with black dots on the $q'$-surface in Fig. 1a (Fig. 1b), respectively 125 (133) sites of the total 169 lattice sites, satisfy the constraint in Eq. (9). For the unoccupied lattice sites this constraint is violated.

These sites correspond to the subsystem's entropic indices with the highly differing values, for instance, $q_1 \approx 1.067$ and $q_2 \approx 1.333$ for $k_1 = 15$ and $k_2 = 3$, respectively. It should be stressed that all $q'$ values were obtained from the linear relation given by Eq. (8). The results for a variable temperature $\beta = 2/(5 - 3q) = 2k/(2k - 3)$ are very similar to the case with $\beta = 1.12$ in Fig. 1b and they are not presented here.

On the other hand, for a simplified model system consisting of two independent subsystems with the number of microstates $W_I$ and $W_{II}$, the exact extended relation, i.e. Eq. (5), can be applied. For simplicity we also assume an equiprobability distribution for subsystem's microstates. Thus, $1/W_I$, $1/W_{II}$ and $1/(W_I W_{II})$ give the probabilities of subsystem's and entire system microstates, respectively. Consequently, Eq. (5) converts into

$$\frac{1 - (W_I)^{1-q_1}}{q_1 - 1} + \frac{1 - (W_{II})^{1-q_2}}{q_2 - 1} = \frac{1 - (W_I W_{II})^{1-q'}}{q' - 1} . \qquad (19)$$

Solving numerically the exact Eq. (19), e.g. for the case $W_I = W_{II} = 2$, one obtains $q' = q'(q_1, q_2)$ values with a high accuracy without using Taylor expansion. For previously chosen pairs of integers $k_1$ and $k_2$ the results are plotted in Fig. 2. Now, 151 sites (black triangles) of the total 169 lattice sites satisfy the constraint given by Eq. (9). We can observe the same tendency: the subsystem's entropic indices with the highly differing values are rejected (see the unoccupied lattice sites in Fig. 2).

Let us focus on another kind of functional dependence, namely $q'$ as a function of all possible probabilities of subsystem's microstates for a fixed pair of indices. First, for two identical subsystems with $W_I = W_{II} = 2$ and equal entropic indices the surface $q' = q'(p_i, p_j; q_1 = q_2 = 1.15)$ calculated by means of Eq. (5) is presented in Fig. 3. The corresponding discrete distributions of microstate probabilities are chosen as $p_i$ and $1 - p_i$ for the first subsystem and $p_j$ and $1 - p_j$ for the second one. As expected, all the $q'$ values are less than 1.15 for every $p_i, p_j \in (0, 1)$. The contour lines, corresponding to constant values on $q'$-surface, are drawn every 0.005 step. They symmetrically spread over the whole basic square $(p_i, p_j)$ around their centre $(p_{i0}, p_{j0}) = (1/2, 1/2)$. The position of the centre corresponds to a maximum value of Tsallis entropy for the composed system. This is in agreement with Beck's viewpoint about the monotonous decrease of entropic index $q(r)$ along the spatial scale $r$, since $q' = q'(p_{i0}, p_{j0})$ attains its minimum value, that can be connected with a maximal spatial scale and thus, with the largest entropy.

In turn, for different entropic indices of two non-identical subsystems, the values of $q' = q'(p_i, p_j; q_1 = 1.08, q_2 = 1.15)$ are expected to be lower with respect to the corresponding previous results and they are, indeed. The results obtained from Eq. (5) are shown in Fig. 4 for $W_I = W_{II} = 2$ and Fig. 5 for $W_I = 3$ and $W_{II} = 2$. For the latter case the discrete distributions of microstate probabilities were chosen as $p_i/2$, $p_i/2$ and $1 - p_i$ for the first subsystem and $p_j$ and $1 - p_j$ for the second one. Now for both figures the role of the general constraint in Eq. (9) becomes relevant. This constraint clearly distinguishes physically admissible solutions for $q'$ among the mathematically correct ones obtained, e.g., from Eq. (5) only. As a result, the size of the domains $(p_i, p_j)$ marked by the external contour line (in both cases corresponding to the smaller value of $q_1$ and $q_2$) is clearly reduced in comparison to the case considered in Fig. 3. Additionally, as a result of different number of microstates ($W_I \neq W_{II}$), a shift of the domain centre to a new position given by ($p_{i0}$, $p_{j0}$) = (2/3, 1/2) can be observed in Fig. 5.

Summarizing, the following relations and constraints were considered: for the $q$-exponential given by Eq. (14) the relation described by Eq. (8) and the general constraint in Eq. (9) (see the black dots in Fig. 1), for a simplified model system with equiprobability distributions the exact relation given by Eq. (19) and the general constraint in Eq. (9) (see the black triangles in Fig. 2). Also, for a simplified model system with assumed discrete probability distributions and the fixed entropic indices the exact relation given by Eq. (5) together with the general constraint in Eq. (9) were involved (see the contour lines in Figs. 3-5).

## 4. Conclusions

The quasi-additive behaviour of $q$-entropy for the case of a system composed of two statistically independent non-identical subsystems is advanced. On the basis of the extended quasi-additivity property of the Tsallis entropies a linear relation between $q_1$, $q_2$ and $q'$ given by Eq. (8) has been obtained. Also a formula for a scaling law in $r$ has been extended by Eqs. (12-13). The postulated general constraint in Eq. (9) yields some limitations (shown in our examples) for values of entropic indices $q_1$, $q_2$ and $q'$ allowed by Eq. (5) only. The author believes that the approach to the case of non-identical subsystems is consistent with the Beck's concept of quasi-additivity.

## Appendix

For $q_1$ and $q_2$ not far from unity, the more accurate non-linear relation between $q_1$, $q_2$ and $q'$ can be obtained by expanding Eq. (6) and Eq. (7) up to the second order in $q_1 - 1$, $q_2 - 1$ and $q' - 1$

$$\begin{aligned} q' - 1 &+ \frac{1}{3}(q' - 1)^2 \left[ \Lambda_1 + \Lambda_2 + 3\langle B_i \rangle \Gamma_2 + 3\langle B_j \rangle \Gamma_1 \right] \\ &= (q_1 - 1)\Gamma_1 + (q_2 - 1)\Gamma_2 + \frac{1}{3}(q_1 - 1)^2 \Lambda_1 + \frac{1}{3}(q_2 - 1)^2 \Lambda_2, \end{aligned} \tag{A1}$$

where $\Lambda_1 = \langle B_i^3 \rangle / D$, $\Lambda_2 = \langle B_j^3 \rangle / D$ and $\Gamma_1$, $\Gamma_2$ and $D$ has been already defined in the text.

## Acknowledgements

The author would like to thank Prof. C. Tsallis for valuable suggestions.

## References

[1] C. Tsallis, J. Stat. Phys. 52 (1988) 479.
[2] M. Gell-Mann, C. Tsallis (Eds.), Nonextensive Entropy−Interdisciplinary Applications, Oxford University Press, New York, 2004.
[3] G. Wilk, Z. Wlodarczyk, Phys. Rev. Lett. 84 (2000) 2770.
[4] G. Wilk, Z. Wlodarczyk, Chaos, Solitons and Fractals 13 (2002) 581.
[5] C. Beck, Phys. Lett. A 287 (2001) 240.
[6] C. Beck, Phys. Rev. Lett. 87 (2001) 180601.
[7] C. Beck, Europhys. Lett. 57 (2002) 329.
[8] C. Beck, G.S. Lewis, H.L. Swinney, Phys. Rev. E 63 (2001) 035303(R).
[9] C. Beck, F. Schlögl, Thermodynamics of Chaotic Systems, Cambridge University Press, Cambridge, 1993.
[10] C. Tsallis, in Complexity, Metastability and Nonextensivity, C. Beck, G. Benedek, A. Rapisarda, C. Tsallis (Eds.), World Scientific, Singapore, 2005, in press [cond-mat/0409631].
[11] Y. Sato, C. Tsallis, in Proc. Summer School and Conference on Complexity in Science and Society (Patras and Olympia, 14-26 July, 2004), T. Bountis (Ed.), Internat. J. of Bifurcation and Chaos (2005), in press [cond-mat/0411073].
[12] C. Tsallis, M. Gell-Mann, Y. Sato, [cond-mat/0502274].
[13] C. Beck, Physica A 306 (2002) 189.
[14] C. Beck, Physica A 277 (2000) 115.
[15] I.S. Gradshteyn, I.M. Ryzhik, Table of Integrals, Series and Products, Academic Press, New York, 1965.

**Figure captions**

Fig. 1. Entropic index $q'$ of the entire system for the $q$-exponential given by Eq. (14) as a function of chosen integers $k_1 = 1/(q_1 - 1)$ and $k_2 = 1/(q_2 - 1)$ calculated by means of approximated relation given by Eq. (8): (a) for $\beta = 2.0$, (b) for $\beta = 1.12$. The black dots symbols correspond to such pairs of entropic indices $q_1$ and $q_2$ that obey the general constraint in Eq. (9).

Fig. 2. Similarly to Fig. 1 but for a simplified model system with equiprobability distributions. The calculations were done with using the exact formula given by Eq. (19). The black triangles symbols correspond to such pairs of entropic indices $q_1$ and $q_2$ that satisfy the general constraint in Eq. (9).

Fig. 3. Entropic index $q'$ for a simplified model system (with assumed probability distributions) as a function of all possible microstate probabilities $p_i$ and $p_j$ for identical subsystems with $W_I = W_{II} = 2$ and fixed pair of entropic indices $q_1 = q_2 = 1.15$. The calculations were done with use of Eqs. (5) and (9). The contour lines spread over the whole basic square $(p_i, p_j)$. This is a typical behaviour for the case of identical subsystems.

Fig. 4. Same as in Fig. 3 but for non-identical subsystems with the fixed pair of different entropic indices $q_1 = 1.08$ and $q_2 = 1.15$ and the equal number of subsystem's microstates $W_I = W_{II} = 2$. The calculations were made with use of Eqs. (5) and (9). The external contour line corresponding to the smaller value of $q_1$ and $q_2$ limits the basic domain of allowed solutions for $q'$.

Fig. 5. Same as in Fiq. 4 but the number of subsystem's microstates is different, that is $W_I = 3$ and $W_{II} = 2$. The symmetry of the basic domain of allowed solutions for $q'$ is changed and its centre corresponding to the largest value of Tsallis entropy is clearly shifted to a new position fixed by $(p_{i0}, p_{j0}) = (2/3, 1/2)$.

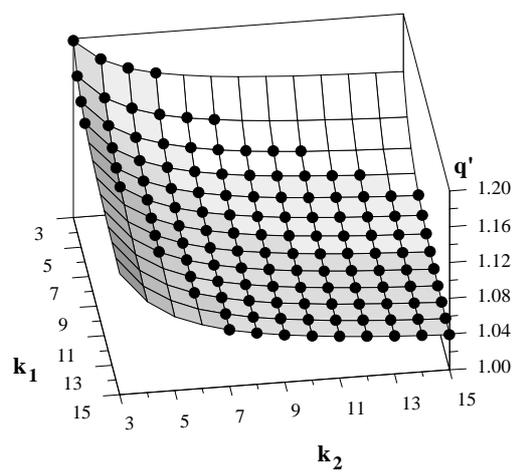

Fig. 1a

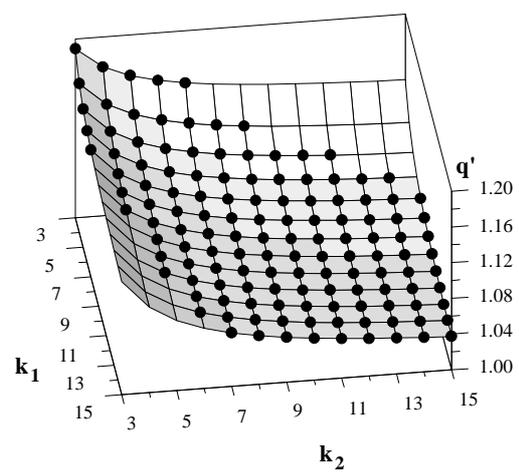

Fig. 1b

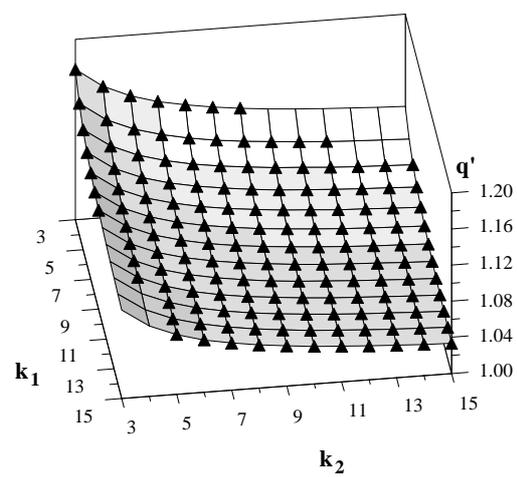

Fig. 2

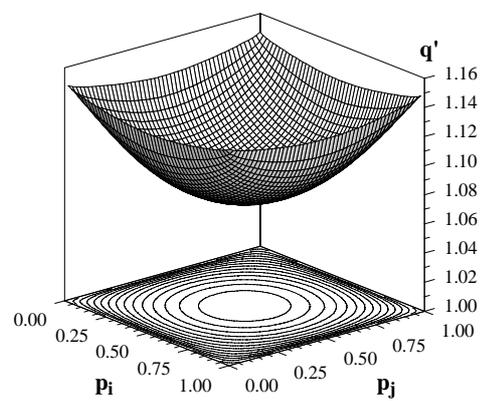

Fig. 3

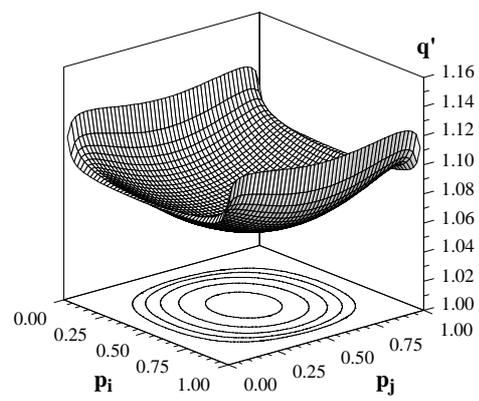

Fig. 4

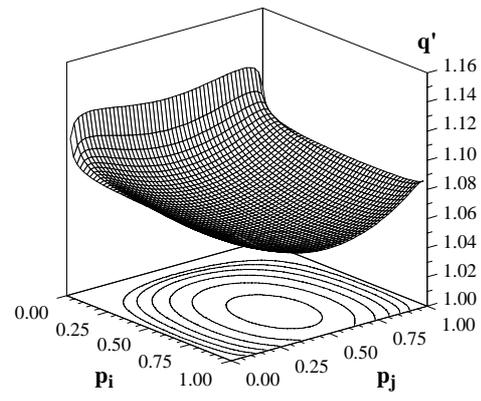

Fig. 5